\pdfoutput=1
\documentclass[%
  prb,%
  letterpaper,%
  onecolumn,%
  floatfix,%
  superscriptaddress%
]{revtex4}

\RequirePackage{graphicx,amsmath,amssymb,bm}
\RequirePackage{hyperref}
\hypersetup{%
  breaklinks = {true},
  citecolor = {blue},
  colorlinks = {true},
  linkcolor = {red},
  pdfauthor = {\textcopyright\ Vitor M. Pereira},
  pdfcreator = {\LaTeX\ and \flqq hyperref\frqq},
  pdffitwindow = {true},
  pdfmenubar = {true},
  pdfpagelayout = {SinglePage},
  pdfstartview = {Fit},
  pdftoolbar = {true},
  plainpages = {false},
}

\newcommand{\Fref}[1]{Fig.~\ref{#1}}

\renewcommand{\eqref}[1]{eq.~(\ref{#1})}

\RequirePackage[usenames,dvipsnames]{xcolor}
\usepackage{soul}

\begin{document}

\title{Conductance signatures of electron confinement induced by strained 
nanobubbles in graphene}

\author{D. A. Bahamon%
\footnote{darioabahamon@gmail.com}%
}
\affiliation{MackGraphe -Graphene and Nano-Materials Research Center, Mackenzie Presbyterian University,
Rua da Consola\c{c}\~{a}o 896, 01302-907, S\~{a}o Paulo, SP, Brazil}

\author{Zenan~Qi%
}
\affiliation{Department of Mechanical Engineering, Boston University, Boston, MA 
02215}

\author{Harold~S.~Park%
}
\affiliation{Department of Mechanical Engineering, Boston University, Boston, MA 
02215}

\author{Vitor~M.~Pereira%
}
\affiliation{Graphene Research Centre \& Department of Physics, National 
University of Singapore, 2 Science Drive 3, Singapore 117542}

\author{David~K.~Campbell%
}
\affiliation{Department of Physics, Boston University, 590 Commonwealth Ave, 
Boston, Massachusetts 02215, USA}

\date{\today}

\begin{abstract}
We investigate the impact of strained nanobubbles on the conductance 
characteristics of graphene nanoribbons using a combined molecular 
dynamics\,--\,tight-binding 
simulation scheme. We describe in detail how the conductance, density of 
states, and current density of zigzag or armchair graphene nanoribbons are 
modified by the presence of a nanobubble. In particular,  we establish that 
low-energy electrons can be confined in the vicinity or within the nanobubbles 
by the delicate interplay between the pseudomagnetic field pattern created by 
the shape of the bubble, mode mixing, and substrate interaction.  The coupling 
between confined evanescent states and propagating modes can be enhanced 
under different clamping conditions, which translates into Fano resonances in 
the conductance traces.
\end{abstract}

%\keywords{graphene, dichroism, graphene nanoribbons, optical absorption, 
%anisotropy}

%\pacs{81.05.ue, 73.63.-b, 77.65.Ly}

% 81.05.ue Carbon-based materials - graphene, 
% 73.63.-b: Electronic transport in nanoscale materials and structures
% 77.65.Ly: Electromechanical effects, strain-induced electromechanical effects, 

\maketitle

% === MAIN MATTER ====================================================

The fine control over nanofabrication techniques has not only increased the 
performance of existing electronic devices~\cite{6186749}, but has also allowed 
the emergence of concept devices based on the strictly quantum-mechanical 
properties of electrons. 
One such proposal is the incorporation of patterned ferromagnetic or 
superconducting films on two dimensional electron gas (2DEG) structures. 
Under the right conditions and design parameters, these can be tailored to
provide non-homogeneous magnetic fields able to interact strongly with the 
underlying electrons in the ballistic transport regime 
\cite{Nogaret,Bending:1990jk,PhysRevLett.72.1518,PhysRevB.48.15166}. 
Ideally, the spatial profile of such fields should be extremely sharp along 
the transport direction and homogeneous in the transverse direction, so that 
the resulting magnetic barrier might behave as an effective momentum filter, 
which is necessary to achieve control of the ballistic 
transmission \cite{PhysRevLett.72.1518,PhysRevB.48.15166}.
In addition, strong and sharp barriers generally beget richer transmission 
characteristics, including the stabilization of confined states within the 
barrier \cite{Xu:2007ty}. 
The same concept has been proposed following the advent of graphene as a 
versatile two-dimensional platform for nanoscale electronic devices, with 
local magnetic barriers being one of several proposed means to confine, guide, 
and control electron flow 
\cite{Martino:2007zl,Masir:2008ve,Ramezani_inHMF,Oroszlany:2008fp,
Kormanyos:2008lq}. The need for robust and tunable confinement strategies is 
more fundamental in graphene electronic devices than in conventional 
semiconductors: on account of their massless Dirac character, charge 
carriers in graphene are vulnerable to the phenomenon of Klein tunneling, and 
cannot be adequately confined by standard electrostatic means, particularly in 
the ballistic regime. 
However, even though the search towards achieving control of the electron flow 
in graphene remains one of the most active research areas when it comes to 
applications of graphene in the electronics industry, little progress has been 
made towards this concept of magnetic confinement. This is partly because 
of the size requirements that call for magnetic barriers that are much smaller 
than the electronic mean free path, and also because of the need to limit the 
spatial extent of the magnetic field within regions equally small, since it might 
be desirable to have portions of the system free of any magnetic fields.
Graphene, with its outstanding electronic and mechanical properties, offers a 
completely new approach towards this goal of local magnetic barriers that can, 
in principle, be modulated on scales of a few angstroms. Owing to the peculiar 
coupling of electrons and lattice deformations, it is possible to perturb the 
electrons in graphene in the same way they would react to an external magnetic 
field by purely mechanical means \cite{Kane:1997,Suzuura:2002}. Several 
authors have envisaged the study of phenomena and applications predicted to 
happen in the presence of magnetic fields by purely mechanical means, exploring 
appropriately engineered strain configurations to achieve desired pseudomagnetic 
field (PMF) profiles \cite{GuineaPRB2008,PereiraPRL2009,VozmedianoPRRSPL2010}. 
The development of Landau quantization in the absence of magnetic fields is one 
such prediction \cite{GuineaNP2010} that was recently confirmed in local 
tunneling spectroscopy experiments \cite{LevySci2010,LuNatureComm2012}. One 
possible application of this ability to create quasi-uniform PMFs over nm 
scales is the fabrication of pseudomagnetic quantum dots \cite{ZenanRTMQD} 
whose sharp resonant tunneling characteristics might provide a very 
sensitive strain detector.
The experiments of Levy \cite{LevySci2010} and Lu \cite{LuNatureComm2012} with 
graphene nanobubbles affirm the potential of strain-engineering for effective 
manipulation of the electronic motion in graphene, and demonstrate the unique 
characteristics of this approach: (i) the ability to generate local PMFs with 
magnitudes that can easily exceed several 100s of Tesla; (ii) the possibility of 
having these fields localized in regions of only a few nm, if strain can be 
locally concentrated; (iii) the prospect of continuously varying the strength of 
the local PMF, in particular being able to establish and remove it on demand; 
and
(iv) not requiring drastic extrinsic modifications of the graphene layer, thus 
preserving most of its intrinsic superlatives, namely the high mobility and the 
Dirac nature of its carriers.
Recently, in order to gain more insight into details of the PMF magnitude 
and spatial profile associated with graphene nanobubbles, as well as to 
understand the role played by typical substrates, the authors with several colleagues 
conducted a 
study of the effects of nano-sized nanobubbles in graphene under different 
geometries and substrate conditions  \cite{QiPRB2014}. 
In order to have a continuous and tunable 
range of deflection, the nanobubbles were generated by inflation under gas 
pressure against selected apertures on the substrate \cite{QiPRB2014}.  
In the present article, 
we revisit this problem from the point of view of electronic transport to 
elucidate the main signatures of circular and triangular nanobubbles, 
and their strong PMF, imprint on the conductance characteristics. 
We are particularly interested in 
whether the large and local PMF leads to scattering and/or confinement that is 
significant enough to translate into modified transmission characteristics.
This point is specially important, regarding the recent observed current 
division in graphene membranes pressurized against triangular 
holes \cite{Bockrath}.
Existing work approaches similar scenarios by straining
graphene according to deformation fields that are
either prescribed analytically or obtained numerically, but
always following from the equations of continuum elasticity
\cite{Gradinar:2013kq,Cosma:2014jk}.
We  tackle this in the same framework developed in 
reference~\citenum{QiPRB2014}, that combines molecular dynamics (MD) and 
tight-binding (TB) calculations. In this approach, the lattice deformation is 
determined fully atomistically for the prescribed substrate and loading 
conditions, and the relaxed atomic positions are used to build a TB description 
of the electron dynamics in the system. The aim is to reduce any bias in the 
description of the electronic system by capturing all the atomic-scale details 
of deformation and curvature, since they play an important role at these scales 
of less than 50\,nm. 
Similar to what is observed for real magnetic barriers~\cite{Xu:2007ty} or 
Gaussian bump deformations \cite{DFariaGD,Settnes:2015kq}, the conductance 
of either zigzag (ZZ) or armchair (AC) graphene nanoribbons (GNR) develops 
marked dips (anti-resonances) at the edge of each conductance plateau. 
We show that this is due to scattering of propagating modes into 
evanescent states confined in the nanobubble. 
The coupling between the confined evanescent state and the 
propagating modes can be enhanced under different clamping and substrate 
conditions, leading to Fano resonances~\cite{UFano,FanoResRMP,Nockel} in the 
conductance traces. Our calculations show that these signatures of 
electronic confinement in graphene nanobubbles are a robust effect, being 
observed irrespective of the orientation of the underlying graphene lattice, for 
circular and triangular graphene nanobubbles on hexagonal boron nitride.

\section{Model and methodology}\label{sec:methodology}

To reproduce the deformation of graphene and its derived transport properties 
as accurately as possible, we implemented a combined MD-TB simulation. 
Molecular dynamics provides the spatial location of the carbon atoms when 
graphene is subjected to gas pressure and a nanobulge forms through the 
substrate aperture. Once the coordinates of each atom are known, the 
nearest-neighbor TB parameters are calculated throughout the system and the TB 
Hamiltonian for the deformed system is built. This Hamiltonian constitutes the 
basis for the calculation of all the local spectral and transport properties. 
Electronic transport is addressed via the lattice representation of the 
non-equilibrium Green's function (NEGF).

It is beneficial to underline from the outset the role we attribute to the 
substrate in our modeling with regards to the electronic structure and 
transport: all the electronic action is taken to happen within the 
graphene sheet, which we assume not to be chemically perturbed in a significant 
way by the presence of the substrate underneath. This amounts to assuming that 
the electronic properties of graphene are completely decoupled from those of 
the substrate, the latter playing a rather passive role from this perspective, 
in that it simply stabilizes the static lattice configuration of graphene on 
which all the electronic action unfolds. This is a reasonable assumption for 
most current experimental scenarios, where graphene is physically transferred 
and deposited on a target substrate with a random orientation of the respective 
lattice directions; it also implicitly assumes substrates without 
reactive/dangling bonds that could strongly interact with those $p_z$ orbitals 
that happen to be in registry and become a significant source 
of disorder. The most important aspect of this scenario of weak electronic 
coupling between graphene and the substrate is that we consider electronic 
conduction taking place \emph{only} through the graphene system, and its 
characteristics are determined solely by the electronic states derived from the 
$p_z$ orbitals in the deformed and curved graphene. This is done so that the 
computations can be easily extended to tens of thousands of atoms, and relies 
on a tight-binding parameterization of the electronic dynamics that has been 
repeatedly shown to be reliable to describe low energy processes such as those 
involved in the electronic conduction.
Moreover, a full \emph{ab-initio} consideration of the 
relaxation, electronic structure and quantum transport is unattainable in this 
context because (i) the deformation fields are highly non-uniform, (ii) 
graphene, substrate and gas atoms have to be all taken into account, and (iii) 
we wish to tackle the characteristic deformation scales seen in the experiments 
quoted above, all of which entail a large number of atoms in the minimal 
supercell. This justifies and motivates the multi-scale approach to this problem 
that we now describe in more detail.

%---------------------------------------------------------------------
% FIGURE 1
%---------------------------------------------------------------------
%
\begin{figure}[h]
\centering
  \includegraphics[scale=0.4]{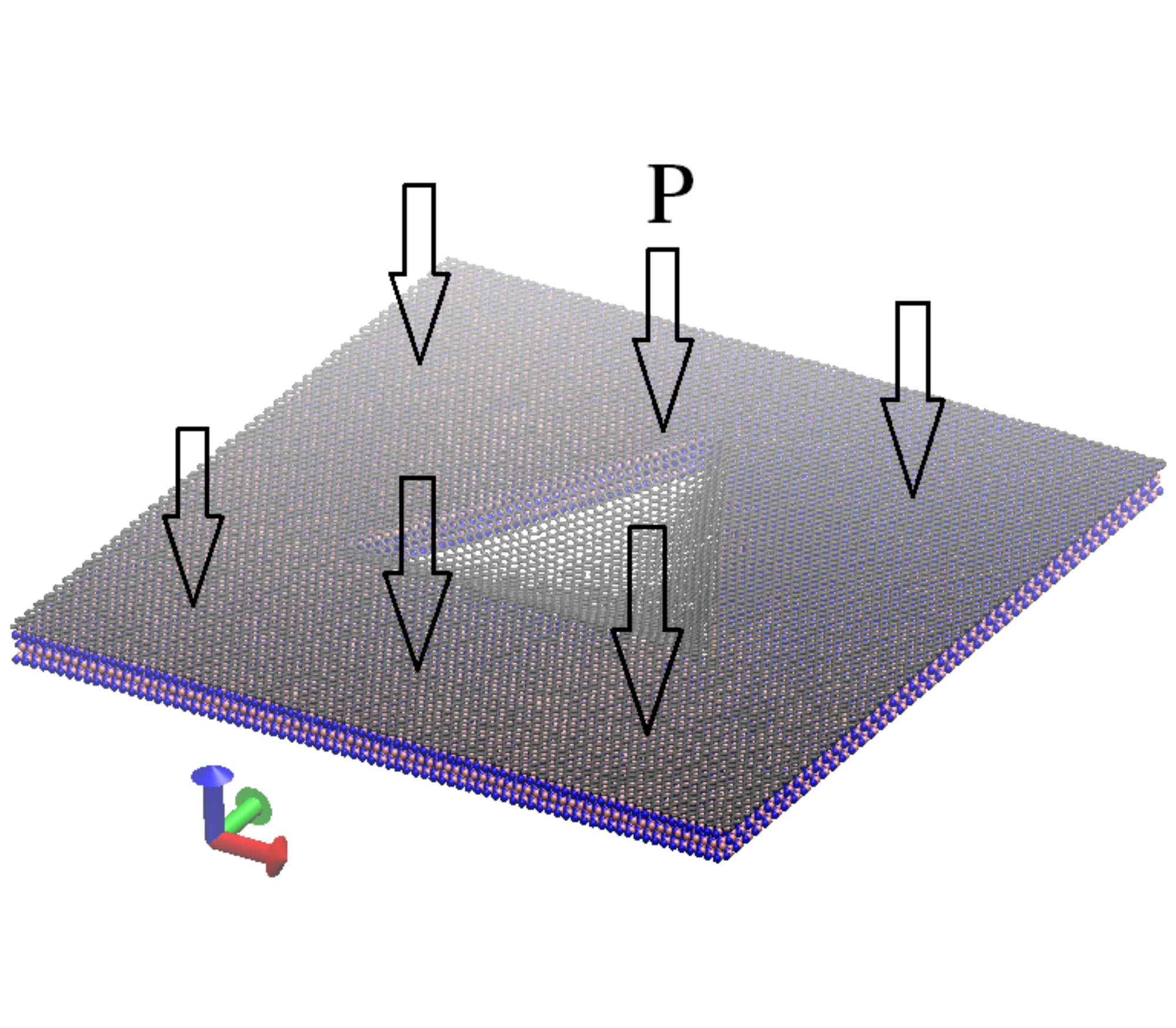}
  \caption{Illustration of a MD simulation cell conveying the 
strategy used to generate the graphene nanobubbles. An aperture 
(a triangle in this case) is perforated on hexagonal boron nitride
on which rests a monolayer of graphene (gray). Argon gas is then 
pressurized against graphene which bulges through the aperture, with 
a deflection that is controlled by the gas pressure. For ease of 
visualization the gas molecules are not shown in the picture above.
Visualization is performed using VMD~\cite{Humphrey1996}.}
\label{fig:schematic}
\end{figure}
%---------------------------------------------------------------------

\subsection{Molecular dynamics simulations}
For an unbiased analysis of the local profile of deformations, the 
mechanical response of the system was simulated by MD with the 
Sandia-developed open source code LAMMPS~\cite{plimptonLAMMPS,PlimptonJCP1995}. 
The MD simulation system consisted of three subsystems: a graphene monolayer, 
a rigid substrate with a central aperture, and argon gas that was used to 
inflate graphene through the aperture to generate a nanobubble. 
An illustration of the system is shown in \Fref{fig:schematic}. 
The Tersoff potential was 
used to describe the C-B-N interactions. The parameters were adopted 
from references \citenum{Sevik:2011kq,Sevik:2012sf,Lindsay:2010zl},
the dimension of the simulation box was 
20$\times$20$\times$8\,nm$^3$ and circular and triangular apertures
were ``etched'' in the center 
of the substrate to allow the graphene membrane to bulge downwards due to the 
gas pressure. In each simulation, the system was initially relaxed for 50\,ps 
before slowly raising the pressure to the desired target by decreasing the 
volume of the gas chamber. Upon reaching the target pressure, the system 
was allowed to relax for 10\,ps, after which deformed configurations were 
obtained by averaging the coordinates during equilibrium. 
Target pressures are determined to yield 
a deflection of 1\,nm. All simulations in the presence of the gas were carried 
out at room temperature (300\,K) using the Nose-Hoover 
thermostat~\cite{hooverPRA1985}. 

Since a previous study established that the magnitudes and space dependence of 
the strain-induced PMFs can be very sensitive to the clamping conditions and
substrate type, \cite{QiPRB2014}, we considered two scenarios to analyze 
how these effects impact the transport signatures. In one case the MD 
simulations are done with clamped boundary conditions, i.e., an ideal system 
consisting only of Ar gas and graphene, and where all carbon atoms outside the 
aperture region were strictly fixed. This is to study the effect of aperture 
geometry without considering the substrate, and is similar to the approach used 
in previous work~\cite{WangJAM2013,GuineaPRB2008,KimPRB2011}. In the second 
scenario, we included a 1\,nm thick substrate of h-BN and its 
interaction with the graphene sheet is explicitly taken into account.
The Ar-BN (gas-substrate) 
interactions were neglected, and the substrate layer remained static during the 
simulation. Most of the graphene layer was unconstrained, except for a 0.5\,nm 
region around the outer edges of the simulation box where it remained pinned.

The choice of the substrate is motivated by is the experimental observation 
that, for certain substrates such as boron nitride, graphene develops a
nonuniform strain strong enough to induce an energy gap $\simeq{20}$\,meV 
at the Dirac point \cite{Hunt21062013,Jung:2015kq,Neek-Amal,San-Jose:2014vn},
to introduce satellite Dirac points \cite{Cloning-Geim,Cloning-LeRoy}, and to 
allow the observation of a Hofstadter spectrum \cite{Hofstadter:1976} in the 
presence of a magnetic field \cite{Hofstadter-Geim}.
Our goal is to assess whether any features in the conductance of the system 
when deformed under realistic conditions of contact with a substrate are 
robust, or dependent on the degree of substrate-graphene interaction. 

\subsection{Tight-binding calculations}

The scattering region used in the electronic transport calculations contains 
the entire MD simulation cell (including the flat portions between the bubble's 
perimeter and the edge of the cell). The cell accommodates 15088 lattice sites, 
an example of which is shown in \Fref{fig:schematic}. For convenience, we take 
the $x$ axis parallel to the ZZ direction. Most low energy electronic 
properties of graphene are captured by the $\pi$ band nearest-neighbor TB 
Hamiltonian
%
%%%% equation TB
\begin{equation}
\label{eq:H}
H =   \sum_{<i,j>}t_{ij} (c_{i}^{\dagger} c_{j} +
c_{j}^{\dagger} c_{i})
,
\end{equation}
%%%%%%%%%%%%%%%%%%%
%
where $c_{i}$ represents the annihilation operator on site $i$ and $t_{ij}$ is 
the hopping amplitude between nearest neighbor $\pi$ orbitals (in the 
unstrained lattice $t_{ij}=t_0 \approx -2.7$\,eV). The link between the MD 
simulation and the TB Hamiltonian is performed when the positions of the carbon 
atoms in the deformed configuration, obtained by MD, are incorporated into the 
TB Hamiltonian through the modification of the hopping parameter $t_{ij}$ 
between all nearest-neighbors. The modification that accounts simultaneously 
for the changing distance $d$ between neighbors and the local rotation of the 
$p_z$ orbitals is given by:
%
%%%% equation tij
\begin{multline}
t_{ij}(d)=V_{pp\pi}(d_{ij})\, \hat{n}_i \cdot \hat{n}_j \\
+ \Bigl[V_{pp\sigma}(d_{ij})-V_{pp\pi}(d_{ij})\Bigr] \frac{(\hat{n}_i \cdot 
\vec{d}_{ij})
(\hat{n}_j \cdot \vec{d}_{ij}) }{d^2_{ij}}
,
\label{eq:tij}
\end{multline}
where $\hat{n}_i$ is the unit normal to the surface at site $i$, 
$\vec{d}_{ij}$ is the distance vector connecting two sites $i$ and $j$, 
and $V_{pp\sigma}$(d) and $V_{pp\pi}$(d) are the Slater-Koster bond integrals 
for $\sigma$ and $\pi$ bonds. Their dependence on the inter-atom distance is 
taken as \cite{PereiraPRB2009,QiPRB2014}
%
%%% equation Vpp
\begin{align}
\label{eq:SKp1}
  V_{pp\pi}(d_{ij}) & = t \, e^{-\beta(d_{ij}/a-1)}, \\
  \label{eq:SKp2}
  V_{pp\sigma}(d_{ij}) & = 1.7 \, V_{pp\pi}(d_{ij}),
\end{align}
where $t=2.7$\,eV, $a\simeq 1.42$\,\AA\ represents the equilibrium bond length 
in graphene, and $\beta = 3.37$ captures the exponential decrease in the 
hopping with interatomic distance. Once the values of $t_{ij}$ are obtained, we 
use the TB Hamiltonian of the 
strained system as the scattering central region, to which two ideal contacts 
are attached. Since the edges of the system are of ZZ or AC type, the central 
region is seamlessly stitched to the contacts resulting in a perfect ZZ or AC 
ribbon. We then study the quantum transport characteristics of such a 
GNR containing a central region deformed by the presence of the nano-bubble.
The zigzag graphene nanoribbon (ZGNR) is created attaching two pristine 
semi-infinite ZZ nanoribbons to the left and right edges of the strained  
graphene square. The metallic armchair graphene nanoribbon (AGNR) is constructed 
by connecting two perfect metallic semi-infinite AGNR to the upper and lower 
edges of the central region.
The conductance of these nanoribbons is calculated within the 
Landauer-B\"uttiker formalisim using Caroli's formula \cite{Caroli,Datta,Jauho}: 
$G=\frac{2e^2}{h} \text{Tr}[\Gamma_qG^r\Gamma_pG^a]$,
where $G^r=[G^a]^{\dagger}=[E+i\eta-H-\Sigma_p-\Sigma_q]^{-1}$ is the 
retarded [advanced] Green's function, the coupling between the contacts and the 
central region is represented by $\Gamma_q=i[\Sigma_q-\Sigma_q^{\dagger}]$, and 
$\Sigma_q$ is the self-energy of contact $q$ which is calculated recursively for 
ZZ and AC contacts~\cite{LopezSancho}. Having calculated the retarded 
and advanced Green's functions, other electronic properties such as the
density of states (LDOS), $\rho_{ii}=-\text{Im}[G^r(\vec{r_i},\vec{r_i},E)]/ 
\pi $, and the total density of states (DOS), $\rho = \text{Tr}(\rho_{ii}) $ 
are readily calculated. 
For a local mapping of the current distribution in the central region we 
consider the current density between nearest neighbors \cite{Caroli},  
$I_{ij}=\frac{2e}{h}\int dE[t_{ji}G_{ij}^<-t_{ij}G_{ji}^<]$, that is calculated 
from the lesser Green's function, and which can be obtained exactly in the 
absence of electronic interactions as \cite{Jauho} $G^< = 
G^r(E)[\Gamma_L(E)f_L(E) + \Gamma_R(E)f_R(E)]G^a(E)$. 
We stress again that the interaction graphene-substrate is included in the MD 
simulation part to realistically describe the interaction and sliding of 
graphene in contact with the substrate by the combined action of gas pressure 
and substrate aperture \cite{nl4007112}. 
From the electronic point of view, the substrate 
plays no direct role in electronic tunneling or other electronic processes.

In order to compare the local current distribution to the spatial pattern of 
the PMF the latter is calculated directly from $t_{ij}$ 
introduced in \eqref{eq:tij} via 
\begin{equation}
  A_{x}(\textbf{r}) - i A_{y}(\textbf{r}) = \frac{2\hbar}{3 t a e}
  \sum_{\textbf{n}} \delta t_{\textbf{r},\textbf{r}+\textbf{n}}
  \, e^{i\textbf{K}\cdot\textbf{n}}
  .
  \label{eq:Adef-full}
\end{equation}
This defines the two-dimensional pseudomagnetic vector potential, 
$\textbf{A}=(A_x,\,A_y)$ \cite{Kane:1997,Suzuura:2002}, from where the PMF is 
calculated using $B=\partial_x A_y - \partial_y A_x$.

\section{Pseudomagnetic fields, mode mixing and confinement}

In order to recognize the  incremental contributions of the different 
factors determining the conductance characteristics of the system (geometry, 
substrate interaction, and edge type of the GNR), we start with the 
simplest scenario described above: a ZGNR where all carbon atoms outside the 
aperture are rigidly (thus artificially) attached to their original position; 
any deformation occurs only within the aperture region under the gas pressure. 
Under this scheme the nanobubble in the middle of the ZGNR is the only 
extended scattering center, which allows us to isolate the effect of the bubble 
geometry and the corresponding PMF on the conductance. 
We chose two representative cases of aperture geometry for discussion: 
triangular and circular. The triangular aperture is particular because it 
begets a PMF that is appreciably uniform within most of the bubble area, and 
which does not alternate in sign within. The circular hole, on the other hand, 
is used because it captures most of the qualitative features of the PMF that 
sets in for a class of different shapes \cite{QiPRB2014}
For a meaningful comparison, circular and triangular bubbles are chosen with 
approximately the same area $\simeq{50}\,\text{nm}^2$, and centered within the 
square simulation cell; specifically, the radius of the circular aperture is 
4\,nm and the side length of the triangle is 10.6\,nm.
In a second stage, we analyze the conductance traces arising from the 
nanobubbles inflated against a h-BN substrate to determine whether the 
graphene-substrate interaction perturbs the conductance traces of the ideal 
clamped situation.

\subsection{Clamped bubbles} \label{sec:clamped}

%---------------------------------------------------------------------
% FIGURE 2
%--------------------------------------------------------------------- 
\begin{figure}[h]
\centering
   \includegraphics[scale=1.1]{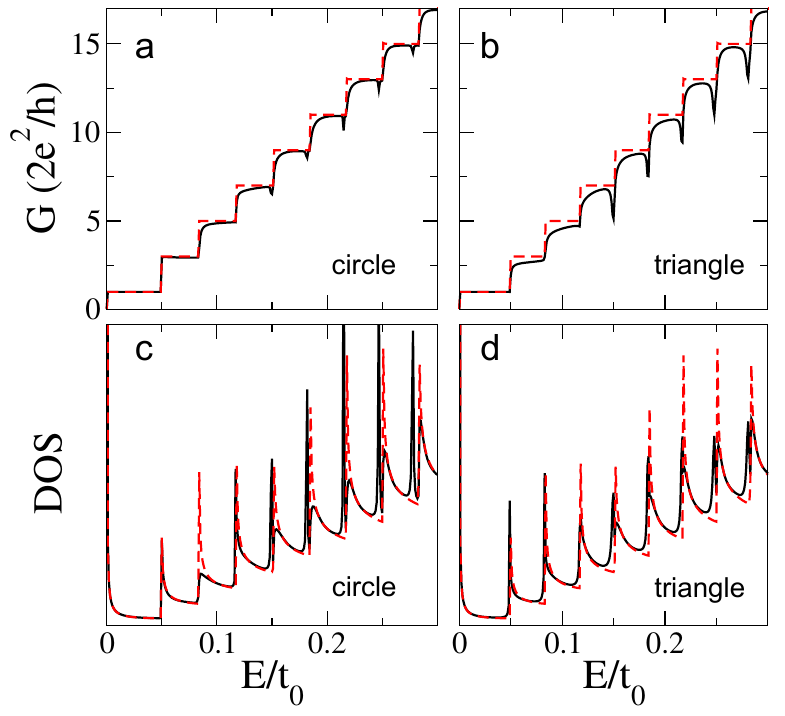}
  \caption{Top rows present the conductance as function of the Fermi energy, 
 $E_F$, for a ZZ GNR 20\,nm wide with an embedded (a) circular and (b) triangular 
nanobubble. Bottom rows correspond to the Density of States (DOS) of the same
ZGNR with (c) circular and (d) triangular bubbles. In all  panels the carbon atoms 
outside the bubble region are rigidly clamped to the substrate and remain 
(artificially) undisplaced. The red dashed lines correspond 
to the conductance and DOS of a pristine graphene ribbon with ZZ edges
state.}
\label{fig:Gclamped}
\end{figure}
%---------------------------------------------------------------------

There is one key feature in the quantum transport of these systems stemming 
from the presence of the central bubble in an otherwise perfect GNR, and which 
is independent of the bubble geometry. Irrespective of the shape, the 
conductance of a ZGNR with $W\simeq{20}$\,nm of transverse dimension with an 
embedded bubble exhibits reproducible dips just at the onset of a every new 
conductance plateau. The conductance traces for circular and triangular 
nanobubbles are shown 
in \Fref{fig:Gclamped}(a)-(b) for a gas pressure of 19\,Kbar, equivalent to a 
deflection of 1\,nm.
The difference in sharpness and depth of these dips, as well as the roundness of 
the conductance steps, can be attributed to the geometry of the bubbles 
which, together with the spatial extent and magnitude of the local PMF, 
contributes to defining the strength of the scatterer.
The weaker the scatterer, the narrower the line-shape of the conductance 
dips will be \cite{Bagwell,Chu:1989kq}. 
The red dashed traces in \Fref{fig:Gclamped}(a)-(b) represent the conductance of the 
ideal ZGNR. 
By direct inspection, we see that the conductance is generally lowered relative 
to perfect quantization, and dips remain sharp for the circular bubble.
The triangular bubble exhibits  larger
reduction from the quantized value within each plateau, together with broader 
dips (notwithstanding, the original plateau structure is still identifiable). 
The spectral fingerprint  of the conductance dips is the appearance of strong 
and narrow peaks in the DOS of the ribbons, just below the van~Hove 
singularities (VHS) of the unpressurised system, as observed in 
\Fref{fig:Gclamped}(c)-(d)

\begin{figure}[h]
\centering
\includegraphics[scale=0.7]{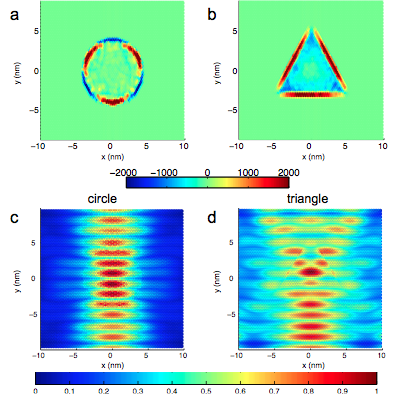}
\caption{Spatial maps of the PMF in the central scattering cell used in the transport 
calculations for the representative cases of a circular (a) and a triangular (b)
nanobulge. The PMF calculated according to \eqref{eq:Adef-full} 
includes the hopping perturbations brought in by bond stretching and bending, 
as per \eqref{eq:tij}. Normalized local DOS for (c) circular and (d) triangular 
bubbles at $E=0.215\,t_0$. 
} 
\label{fig:PMFclamped}
\end{figure}

Before proceeding further with our analysis we want to discuss the origin and 
physics behind the shallow and sharp features observed right before the onset 
of the plateaus (in the conductance) or the VHS (in the DOS). This resonant 
behavior is a multimode effect previously observed in quasi-one dimensional 
systems with impurities \cite{Faist:1990jk,Kander:1990sf}, finite-range
local potential scattering \cite{Vargiamidis:2005mz,PhysRevB.60.10962}, and 
short-range impurity potentials 
\cite{Bagwell,Chu:1989kq,Gurvitz:1993vn,PhysRevB.61.5632,
PhysRevB.70.245308}. It can be understood by recalling that in quantum 
wires electric current is carried by independent transverse modes. When an 
impurity is present an electron incident upon the defect in a given mode will 
be scattered into a number of available modes with the same energy, 
including evanescent states \cite{Bagwell}. The transition probability for this 
process depends on the density of final modes and, therefore, by virtue of the 
high density of evanescent states at the edge of each sub-band (mode), the 
electron has a high probability of scattering to an evanescent state, which is a 
state predominantly confined within the defect region, with an energy close to 
the bottom of the sub-band \cite{Bagwell}. 
Of course, the transition rate depends also on the scattering potential itself, 
in addition to the density of evanescent states. As we outlined in the 
introduction, electrons in graphene perceive non-uniform local changes in the 
electronic hopping parameter as a PMF, and it is this non-uniform PMF pattern 
created by the inflation of graphene that determines the strength of the 
scattering at each nanobubble. The detailed analysis of the PMF created 
by the clamped circular and triangular nanobubbles, and other geometries
not considered here, can be found in reference \citenum{QiPRB2014}. For our 
current purposes, \Fref{fig:PMFclamped}(a)-(b) shows the spatial profile of 
the PMF in the two geometries considered.
We briefly recall that one of the leading characteristics of the PMF 
distribution arising from an inflated nanobubble is an intense magnetic barrier 
that is narrowly localized within a few atomic distances from its perimeter. 
This results from the large bending and high bond stretching that occurs at the 
edge of the apertures. Different geometries have an impact in the local polarity 
of the PMF and its magnitude and space dependence in the central regions of the 
bubble. 

%---------------------------------------------------------------------
% FIGURE PMF CLAMPED (4)
%---------------------------------------------------------------------
%
%---------------------------------------------------------------------

The PMF graphs in \Fref{fig:PMFclamped}(a)-(b) show that the circular bubble 
has high PMF barriers ($\sim{2000}$\,T) at the perimeter, followed by a rapid
decay towards the center of the bubble. 
Triangular bubbles, on the other hand, create PMFs of magnitude equally large 
around the perimeter and a roughly constant field of $\sim{100}$\,T in the inner 
central area. Unlike the circular one, in triangular nanobubbles the intensity 
and polarity of the peripheral barrier remains constant at all the three edges.
Based on the this, we can attribute the conductance dips observed in 
\Fref{fig:Gclamped} to scattering of propagating modes into a confined state 
around the bubble. However, it remains unclear how the wave function of the 
confined electron is distributed under such different strengths and 
patterns of PMF created by the bubbles.
To clarify this point, let us inspect the LDOS maps shown in 
\Fref{fig:PMFclamped}(c)-(d), each taken at the energy of the conductance 
dips observed at $E=0.215\,t_0$. We see no fingerprint of a strictly 
confined state: the shape of the bubble itself is not even identifiable
in either panel and, although the highest values of the LDOS are found within 
the bubble region, they are not significantly different from those outside. 

To interpret these maps it is important to note that the unpressurized 
conductance of these systems at $E=0.215\,t_0$ is
$G(0.216\,t_0) = (2e^2/h)\times 11$. From the conductance 
quantization sequence of an ideal GNR, $ G = (2e^2/h) (2n +1) $ 
\cite{PeresCQ,PhysRevB.78.161409}, we conclude that there are 5 conducting 
modes in an ideal GNR at the energy represented in \Fref{fig:PMFclamped}(c)-(d).
The inclusion of the bubbles brings only a small change to this tally, 
as \Fref{fig:Gclamped} shows that the conductance in their presence is, 
for the most part, scarcely modified: at $E=0.215\,t_0$ one or more channels are 
backscattered because $G=(2e^2/h)\times(10.1)$ for the circle, and $G= 
(2e^2/h)\times(9.4)$ for the triangle. 
Hence, despite the nominal suppression of 1 to 2 conducting modes, the 
conductance is never zero at these energies and, consequently, the LDOS maps 
include contributions from conducting, backscattered, and confined states in the 
same picture. 

%---------------------------------------------------------------------
% FIGURE IMAP CLAMPED (6)
%---------------------------------------------------------------------
%
\begin{figure}[h]
\centering
\includegraphics[scale=1]{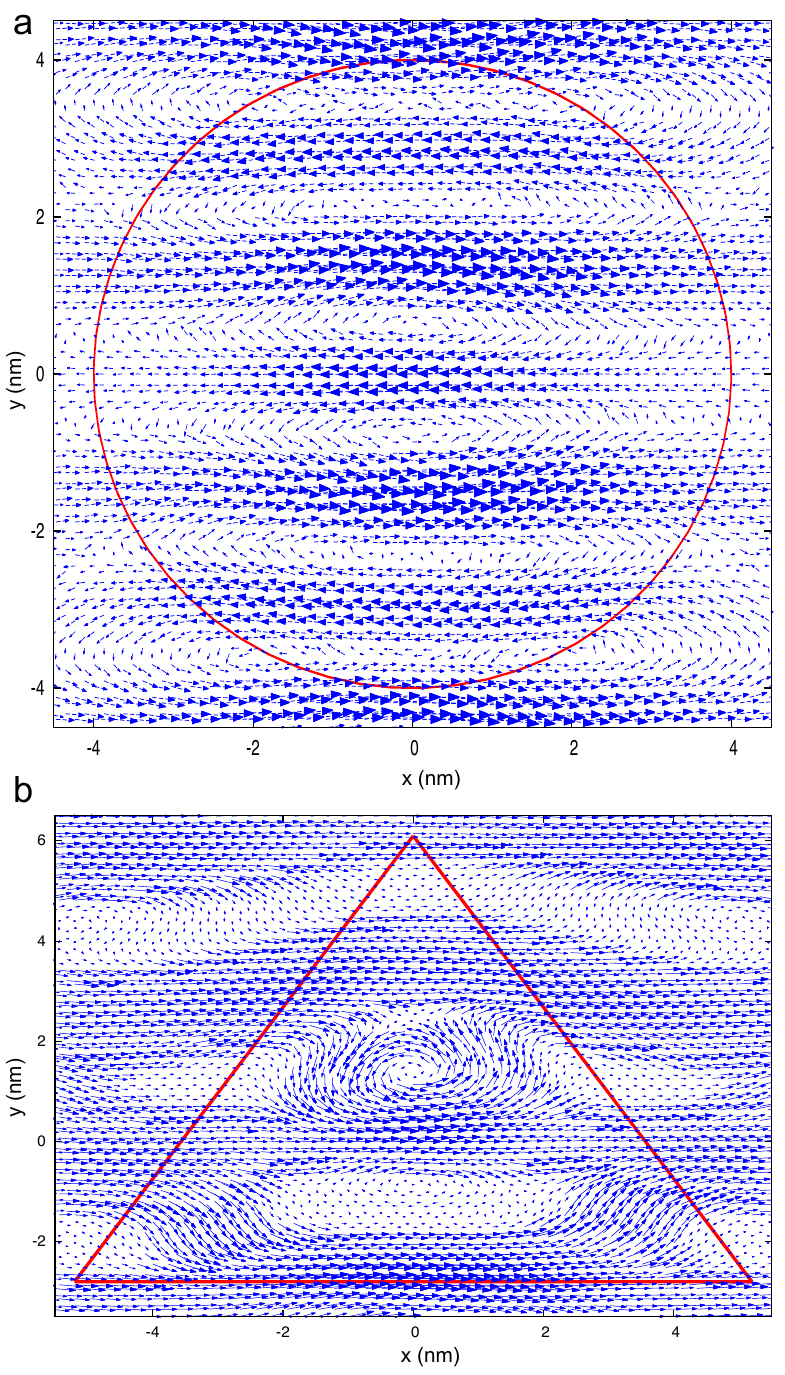} 
\caption{(Color online) 
Current density at $E=0.215\,t_0$ around the clamped circular bubble (a)  
and  triangular bubble (b). The red outline marks the 
portion of the system corresponding to the bubble region. Each blue arrow 
indicates the local current flow, and has a magnitude proportional to the 
current at each lattice site. 
} 
\label{fig:Iclamped}
\end{figure}
%---------------------------------------------------------------------

A better insight into the extent to which the local PMF arising from 
different geometries disrupts the electron flow can be obtained from the 
local current density that we have calculated at each C-C bond as described 
earlier, and whose results are presented in \Fref{fig:Iclamped}. 
The current map shown in \Fref{fig:Iclamped}(a) for the circular bubble reveals 
current streams where the current is directed forwards and backwards in an 
alternating pattern, signaling electron trapping within and its bouncing back 
and forth by the  action of the  strong PMF barriers at the perimeter of the 
bubble (cf. \Fref{fig:PMFclamped}). 
Over the central region of the bubble the current remains 
predominantly horizontal by virtue of the negligible PMF inside a circular 
bubble. These strong bands decay outside bubble, confirming that this current 
pattern is associated with an evanescent mode created by the bubble through mode 
mixing.
Contrarily to the circle, a triangular bubble sustains a high and constant PMF 
$\sim{100}$\,T in the inner central region (cf. \Fref{fig:PMFclamped}). 
Inspection of the current's spatial distribution in \Fref{fig:Iclamped}(b) 
reveals that the PMF within is seemingly enough to permanently trap a fraction 
of the electronic density in closed orbits, as suggested by the presence of a 
local eddy of current of at the center of the bubble.
We note that an electron in graphene with energy $E=0.215\,t_0$ in a constant 
magnetic field of 100\,T has a magnetic length $\ell_B \approx 2.6$\,nm and a 
cyclotron radius of $r_c=\ell_B^2k_F \approx 6.8$\,nm. Since such $r_c$ is 
larger than the bubble, and since other geometries still display conductance 
dips despite the absence of such localized current features, we conclude that 
those effects are not just dominated by the PMF, but bubble geometry and mode 
mixing are important ingredients. 
Finally, note that an electron should have an energy higher than $E \approx 
\hbar v_F \pi / L$ to be sensitive to a scatterer of typical size $L$. The 
average radius of the substrate apertures that we considered is $L \approx 
4$\,nm, which means that only above energies of $E \approx 0.16\,t_0$
should the electrons begin to be noticeably affected by the presence of the 
nanobubble. This estimate is quantitatively consistent with the fact that 
the conductance dips and DOS peaks, observed in  \Fref{fig:Gclamped}, 
only develop above this energy, and are not present at lower energies.

\subsection{Nanobubbles on hexagonal boron nitride substrate}

Whereas the previous section discusses transport in the presence of a 
nanobubble, but having graphene rigidly clamped everywhere except the aperture 
region, in a realistic scenario the graphene-substrate interaction must be
accounted for. The pressure-induced bulging of the graphene sheet through the 
aperture will be accompanied by its sliding and stretching in the regions 
outside the hole. The final strain distribution will thus be different which, in 
turn, will lead to modifications of the PMF barriers. Since the modification of 
electronic conductance discussed above stems from these barriers, one should 
naturally assess how robust they are in a realistic substrate scenario.
To answer this question, we explicitly incorporate the graphene-substrate 
interaction at the atomistic level by carrying out MD simulations of triangular 
and circular graphene nanobubbles on a h-BN substrate, letting all the atoms in 
graphene to relax under the constraint imposed by the gas pressure. 
The PMF that obtains in this case is very similar to that shown previously in 
\Fref{fig:PMFclamped}(a)-(b). This is, of course, not surprising given that 
outside the aperture region graphene is still being pressed against the rigid 
BN substrate; the magnitudes of the fields are, however, smaller, which is a 
direct consequence of the in-plane relaxation of the carbon atoms and the 
smaller in-plane strain that, consequently, sets in for the same deflection 
imposed on the bubble.

The implications of the modified PMF pattern to the conductance can be analyzed 
in two different energy ranges, according to whether the electron's Fermi 
wavelength, $\lambda_F = k / 2\pi$, is larger ($E < 0.150\,t_0$) or smaller 
($E > 0.150\,t_0$) than the characteristic size of the central nanobubble. 
%
%---------------------------------------------------------------------
% FIGURE CONDUCTANCE SUBSTRATE (8)
%---------------------------------------------------------------------
\begin{figure} [h]
\centering
\includegraphics[scale=0.6]{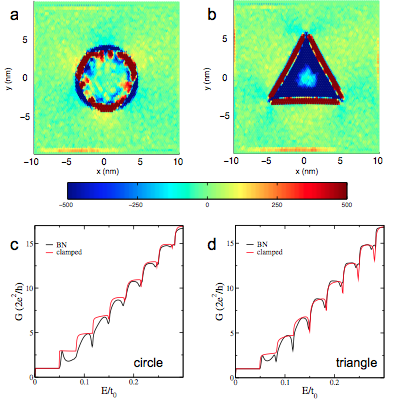}
\caption{(Color online) 
The top row shows the PMF spatial distribution for (a) circular and (b)
triangular bubbles on h-BN substrate.
Bottom rows show the onductance as a function of $E_F$ for ZZ 
nanoribbons 20\,nm wide placed on  
h-BN substrate, and containing a: (a) circular nanobubble and (b) triangular
nanobubble. The red  
lines represent the conductance of the same geometry bubble in the clamped
configuration.}  
\label{fig:GBubble_subs}
\end{figure}
%---------------------------------------------------------------------
%
In \Fref{fig:GBubble_subs}(c)-(d) we show the conductance of a ZGNR with 
embedded circular and triangular bubbles on h-BN; we can see that the 
conductance traces -- specially at low energies -- are now richer than before.
Interestingly, there is no marked difference between the two geometries; 
at higher energies, the presence of the bubble translates only into shallower 
and wider conductance dips.

%---------------------------------------------------------------------
% FIGURE LDOS  SUBSTRATE (9)
%---------------------------------------------------------------------
\begin{figure} [h]
\centering
\includegraphics[scale=0.6]{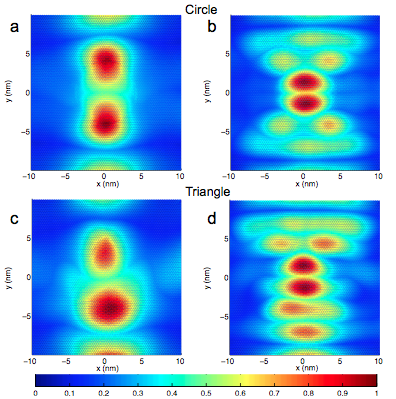}
\caption{(Color online) 
The top row shows the normalized local DOS for a circular bubble in 
graphene lying on a h-BN substrate for (a) conductance peak at 
$E=0.05\,t_0$ and (b) conductance dip at $E=0.116\,t_0$.
The bottom row refers to a triangular bubble on h-BN 
substrate, at (c) the conductance peak for $E=0.05\,t_0$ and (d)
the conductance dip seen at $E=0.116\,t_0$.
} 
\label{fig:LDOSsubs}
\end{figure}
%--------------------------------------------------------------------- 

One new feature detected in \Fref{fig:GBubble_subs}(c)-(d)
is the presence of a resonant peak right at the start of 
the second plateau at $E \simeq{0.05\,t_0}$, and which replaces the 
conductance plateau of the unstrained system. The dips and resonances in the 
conductance are just two particular manifestations of a Fano resonance in the 
electron's scattering cross-section \cite{UFano,FanoResRMP,Nockel} that get 
imprinted in the conductance. In simple terms, a Fano resonance is characterized 
by a transmission probability of the form
\begin{equation}
  T(E) ~ \propto ~ \frac{(\epsilon + q)^2}{\epsilon^2 + 1},
  \qquad
  \epsilon=E-E_{\text{res}}
  \label{eq:Fano}
\end{equation}
in the neighborhood of $E=E_{\text{res}}$, where $\epsilon$ is 
the reduced energy and $q$ the phenomenological Fano asymmetry parameter 
measuring the degree of coupling between a localized (evanescent) state and 
propagating states \cite{FanoResRMP,SataninCAFano}. Whereas in general the 
lineshape described by \eqref{eq:Fano} has a characteristic asymmetric profile, 
if the coupling is strong ($|q| \rightarrow \infty$) it reduces to a resonant 
symmetric peak (Breit-Wigner), while weak coupling ($|q| \rightarrow 0$) is 
characterized by a a dip, or anti-resonance. 

\begin{figure} [h]
\centering
\includegraphics[scale=1.1]{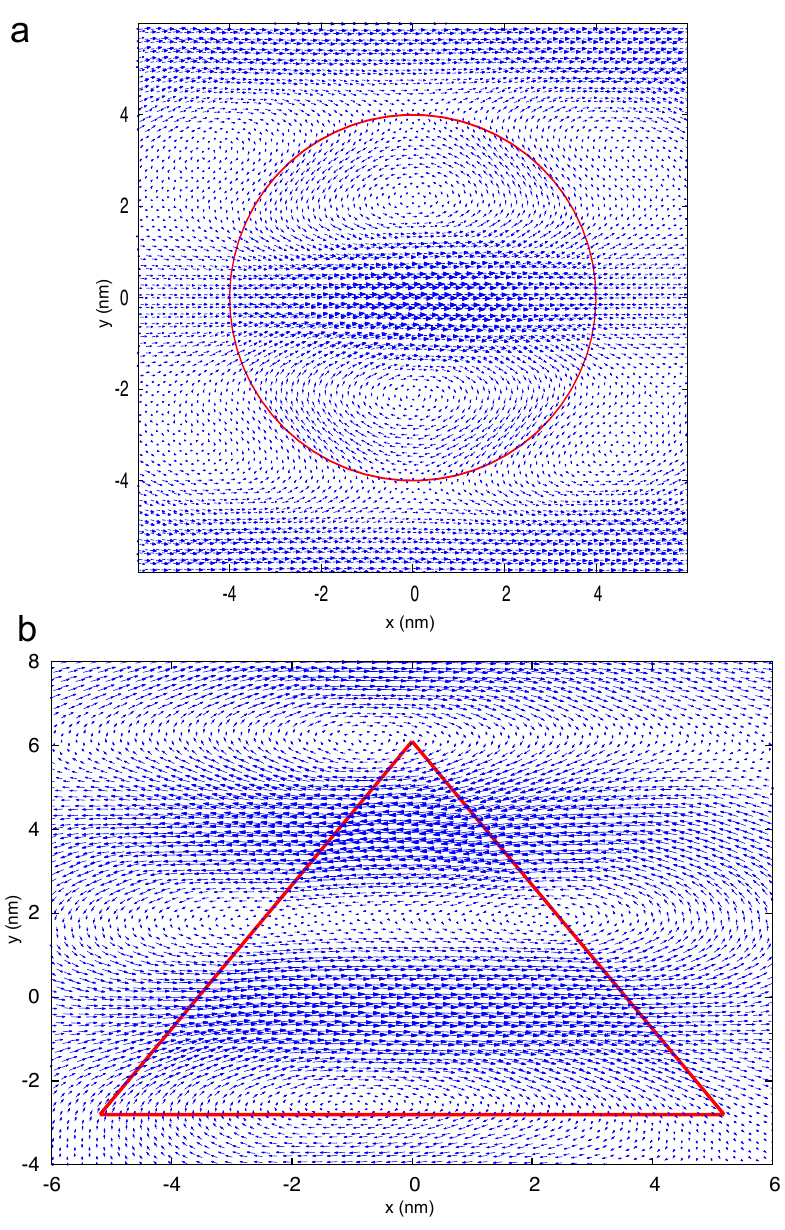}
\caption{(Color online) 
Current density in the vicinity of the (a) circular  and (b) triangular bulges when 
graphene lies on h-BN substrate, both at $E=0.116\,t_0$. The red outline 
marks the portion of the system corresponding to the bubble region.
} 
\label{fig:Isubs}
\end{figure}

To elucidate the origin of the low-energy resonance it is instructive to 
inspect the LDOS at that energy, which is shown in left panels of 
\Fref{fig:LDOSsubs}.
The LDOS in the presence of the circular bubble on h-BN is strongly peaked in 
the regions between the top and bottom edges of the aperture and the 
outer edges of the ribbon. Such an enhancement of the LDOS at the edges 
constitutes a fingerprint of coupling between states \cite{BahamonPRB2010}. 
For this energy $E \simeq{0.05\,t_0}$ at the threshold of the 1st to 2nd 
conductance plateau, the current is carried by a single mode (one can notice 
that $G=G_0$ throughout the 1st plateau) which is strongly localized around the 
edges of the nanoribbons because it is one of the characteristic edge states of 
a ZGNR. 
The LDOS profile in \Fref{fig:LDOSsubs}(a) shows the tendency to localize 
electrons between the perimeter of the circular bubble and the ribbon edges, 
which means that the entire current path coming from the ZZ edge mode overlaps 
spatially with the localized state, leading to a strong-coupling scenario 
between the confined and propagating modes.
This, of course, is a consequence of the underlying PMF for this case: the fact 
that there is a considerable ``leakage'' of the PMF between the aperture and the 
outer edge drives electron confinement in that region of strong field and 
promotes the localization of electrons in a region through which all 
the current would be passing, thus promoting a strong coupling that leads to 
a well defined resonance.
A comparison between panels a and c at the same energy for the triangular 
bubble shows, for the latter, an asymmetric enhancement of the LDOS in the 
vicinity of the upper and lower edges of the ribbon. As a result, the 
coupling to the propagating mode will not be as strong, which explains the fact 
that the resonance at $E \simeq{0.05\,t_0}$ in \Fref{fig:GBubble_subs}(d) is not 
as sharp as it is for the circular bubble.
In contrast to the conductance resonances, the LDOS snapshots associated 
with dips are characterized by a strong enhancement in the central 
area, as can be seen in panels b and d of \Fref{fig:LDOSsubs} for the 
conductance dip at $E= 0.116\,t_0$.
For completeness, we show in \Fref{fig:Isubs} the respective current densities 
at the $E=0.116\,t_0$ dip, which support the previous interpretation, but show 
that the tendency for current localization is diminished in comparison with the 
rigidly clamped scenario, a consequence of the reduced strain in the present 
case.

%
%---------------------------------------------------------------------
% FIGURE I map substrate  (10)
%---------------------------------------------------------------------
%---------------------------------------------------------------------

%---------------------------------------------------------------------
% FIGURE armchair (11)
%---------------------------------------------------------------------
\begin{figure} [h]
\centering
\includegraphics[scale=1.1]{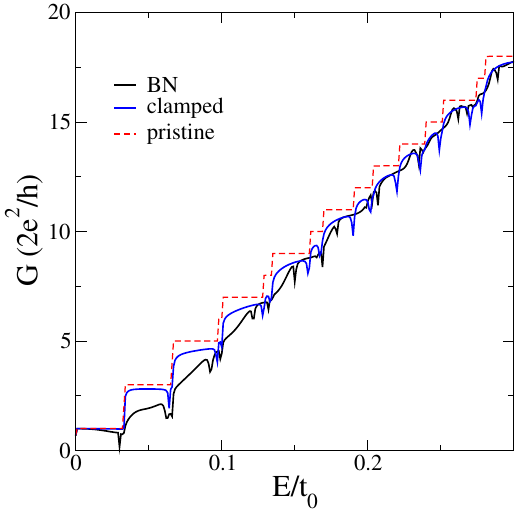}
\caption{(Color online) 
Conductance of a 20\,nm AC nanoribbon with a triangular bubble in the central 
region. The different curves correspond to the conductance of a 
pristine ribbon (red), a bubble under clamped conditions (blue), and a bubble 
on the h-BN (black) substrate. 
}  
\label{fig:armchair}
\end{figure}
%---------------------------------------------------------------------

Finally, we note that the type of graphene lead considered to compute the 
conductance has no bearing on the validity of the discussion and conclusions 
above. To illustrate that, we show in \Fref{fig:armchair} the conductance of 
the same nanobubbles obtained with AC graphene nanoribbons as leads. This was 
done by connecting metallic AC leads to the vertical sides of the square 
system cell.
The resulting conductance profiles are entirely similar to the behavior seen 
in the ZZ transmission configuration, and the differences observed in the 
triangular case are due to the different orientation of the triangle (a 
90$^o$ rotation) with respect to the incoming current. 

\section{Conclusions}

Using a combined molecular dynamics\,--\,tight-binding simulation scheme we 
have investigated the electronic transport properties of graphene 
nanostructures containing circular and triangular nanobubbles, and 
under two graphene-substrate adhesion conditions.
The local strain that develops within and nearby the bubble leads to rich 
patterns of strong PMF with alternating polarity on length scales of a few 
nm. The combination of both strong field and spatially sharp reversal of its 
polarity intuitively suggest a tendency for electron localization at certain 
energies. We have determined how this localization manifests itself (and 
impacts) the electronic transport. Analyses of the LDOS and local current 
distribution reveal the microscopic details of this localization process, and 
establish that low-energy electrons can be confined in the vicinity or within 
the nanobubbles by the interplay of the specific PMF barrier created by the 
geometry of the bubble, mode mixing, and substrate interaction. 
Interestingly, graphene substrate interaction -- unavoidable in real samples -- 
facilitates the appearance of confined states at the same time that it 
determines their coupling to the propagating ones. At low energies, 
the coupling of the evanescent electron states in the vicinity of the 
nanobubbles leads to two distinct signatures in the conductance as a function 
of $E_F$: (i) the appearance of peaks, or Breit-Wigner resonances, when the 
evanescent states spread considerably to the outside of the nanobubble; (ii) 
dips, or anti-resonances, when these states are confined mostly inside the 
nanobubble by the back and forth scattering of electrons between the PMF and, 
consequently, couple less effectively to the continuum.
We conclude that, even though under realistic conditions the interaction 
between graphene and the substrate is seen to modify the magnitude and spatial 
profile of the PMF in relation to an ideal (clamped) scenario \cite{QiPRB2014}, 
there remains a significant tendency for electron confinement under the 
rearranged local strain. 

\section{Acknowledgments}
VMP acknowledges support through Singapore NRF CRP grant ``Novel 2D 
materials with tailored properties: beyond graphene'' (R-144-000-295-281).
ZQ acknowledges the support of the Mechanical Engineering and Physics 
Departments at Boston University.

%---------------------------------------------------------------------
%   References
%---------------------------------------------------------------------
\bibliographystyle{apsrev}
\bibliography{MD_QT}

\end{document}